**Comparable system-level organization of Archaea and Eukaryotes**

J. Podani[1,2], Z. N. Oltvai[1,3], H. Jeong[4], B. Tombor[3], A.-L. Barabási[1,4], and E. Szathmáry[1,2]

[1]Institute for Advanced Study, Collegium Budapest, H-1014 Budapest, Hungary

[2]Department of Plant Taxonomy and Ecology, Eötvös University, H-1117 Budapest, Hungary

[3]Department of Pathology, Northwestern University Medical School, Chicago, IL  60611, U.S.A.

[4]Department of Physics, University of Notre Dame, Notre Dame, IN  46556, U.S.A.

**A central and long-standing issue in evolutionary theory is the origin of biological variation upon which natural selection acts[1]. Previous hypotheses suggested that evolutionary change represents an adaptation to the surrounding environment within the context of inherent constraints derived from the organisms' own internal characteristics[1-3]. Elucidation of the origin and evolutionary relationship of species has been complemented by nucleotide sequence-[4] and gene content[5] analyses, with profound implications for recognizing life's major domains[4]. Yet, our understanding of the evolutionary relationship among species may be further expanded if their systemic higher level organization is also compared. Here, we employ multivariate analytical approaches to evaluate the biochemical reaction pathways characterizing 43 species. The comparison of information transfer pathways demonstrates a close relationship of Archaea and Eukarya. Similarly, although eukaryotic metabolic enzymes have been shown being primarily of bacterial origin[6], the pathway-level organization of archaeal and eukaryotic metabolic networks also prove more closely related. Our analyses therefore reveal a similar systemic organization of Archaea and Eukarya, and suggest that during the symbiotic origin of eukaryotes[7-9] the incorporation of bacterial metabolic enzymes into the proto-archaeal proteome was constrained by the host's preexistent metabolic architecture.**

To begin developing a system level understanding of the evolutionary and organizational relationships among species, we have compared several characteristics of the core metabolic- and information transfer pathways of 43 species representing all three domains of life (i.e., Archaea, Bacteria, and Eukarya), based on data deposited in the WIT integrated-pathway genome database[10]. We have previously established a graph theoretic representation of the biochemical reactions taking place in the metabolic- or information transfer network of a given organism (see ref. 11 and the Supplementary Material for details). With derived matrices in hand, we first created four separate data sets for each organism, comprised of substrates and enzymes of their metabolic- and information transfer networks, respectively. This then allowed us to determine in each data set if a particular substrate or enzyme is present or absent in a given organism, and also to systematically rank order all the substrates and catalyzing enzymes based on the number of biochemical reactions in which they participate. We then compared the individual datasets using several different multivariate analytical approaches, including neighbor joining (NJ: the simplest distance-based cladistic method[12]), unweighted group average clustering (UPGMA: the most commonly used hierarchical classification method[13]), ordinal clustering (OC: hierarchical method suited specifically to rank orders[14,15]) and non-metric multidimensional scaling (NMDS: the most general ordination procedure[15]), providing us insight into various structural aspects of the same database.

First, information pathways (INFO) were examined. Analysis of both substrates (Fig. 1*a, b, e, f, i*) and catalyzing enzymes (Fig. 1*c, d, g, h, j*) provided ordinations (Fig. 1*a-d*), hierarchical classifications (Fig. 1*e-h*) and unrooted trees (Fig. 1*i, j*), and the results of all three approaches proved highly congruent with one another. Regardless whether ordinal information (i.e., rank ordering) (Fig. 1*a, c, e, g*) or mere presence/absence (P/A) (Fig. 1*b, d, f, h-j*) was considered and whether the substrate or enzyme dataset was used, our analyses suggest clear separation of Bacteria (B) both from Archaea (A) and Eukarya (E). This finding is in general agreement with the cladistic results based on ribosomal RNA sequences[4] or gene content[5]. However, there are many fine details of the results that deserve more attention. For instance, even though the sample included many more Bacteria than Eukaryotes and Archaea together, the within-group differences in Bacteria are considerably smaller than within Archaea or Eukarya, e.g. manifesting itself in the ordinations as a very compact scatter of points representing Bacteria (B) (Fig. 1*a-d*). Furthermore, it is evident that Archaea and Eukarya are not merely close to each other, but are essentially inseparable in most cases. For instance, the points representing Archaea and Eukaryotes in both P/A-based ordinations and in the ordinal case for the enzyme-based ordinations form rather elongated point swarms, whose internal cohesion and segregation are therefore not supported (Fig. 1*b-d)*. The only ordination and classification suggesting fair distinction between Archaea and Eukarya are those based on ordinal information from the substrate data (Fig. 1*a, e*).

Next, we compared the systemic organization of metabolic networks. Analyses of data for both substrates (Fig. 2*a, b, e, f, i*) and catalyzing enzymes (Fig. 2*c, d, g, h, j*) of intermediate metabolism provide us with a picture slightly more complex than for the INFO data, but still with many similarities and agreements with the above results. Archaea and Eukarya are clearly recognizable as intact groups in all hierarchies, the exception being the Crenarchae, *A. pernix* (it forms a separate cluster of its own in Fig. 2*f* [arrow]- note, however, the contentious phylogenetic classification of this organism[16]). With the exception of one analysis (Fig. 2*g*) ordinal information (i.e., rank ordering) on substrates (Fig. 2*a, e*) and enzymes (Fig. 2*c, g*) suggests clear separation of Bacteria (B) both from Archaea (A) and Eukarya (E), the latter two being relatively close to each other. When we analyzed the P/A information (Fig. 2*b, d, f, h-j*) the NJ approach on both the substrate or enzyme dataset (Fig. 2*i, j*) and the NMDS approach on the enzyme data (Fig. 2*d*) support a similarly clear separation. Still, the NMDS approach on the substrate data (Fig. 2*b*) and the UPGMA dendrograms on the substrate (Fig. 2*f*) and enzyme data (Fig. 2*h*) indicate a substantially looser association.

Of interest is the clear separation of parasitic bacteria (designated as B2, the rest being B1), including, e.g., *C. pneumoniae, M. genitalium, R. prowazekii, T. pallidum, B. burgdorferi*, that possess an evolutionarily reduced genome[17]. B2 appears as an intact group in all but one trees (Fig. 2*h*). On the other hand, B1 is an intact large group in both trees obtained on the basis of P/A data, whereas OC recognizes two subclusters within B1 (see Fig. 2*e, g*). The NJ method also separates the two groups of Bacteria on the basis of P/A data of both substrates and enzymes (Fig. 2*i, j*). B2 is a loosely arranged group at one end of the unrooted tree, followed by Eukarya and then Archaea. On the other end we find the taxa of B1. Thus, Archaea and Eukarya are inserted between the two bacterial groups in the tree, and it cannot be rooted to produce a cladogram with B1 and B2 as sister groups. It is of course not surprising that parasitic organisms more readily lose metabolic genes, but it is revealing that they do this in a way that statistically put them into the same group, despite their phylogenetic distance.

The analysis and comparison of the systemic, higher level organization of 43 species revealed unanticipated features of the relationship of species within and among the major domains of life. For example, the comparison of systemic attributes of metabolic networks allowed the recognition of convergent evolutionary trends (B2 group) reflected only by the metabolism of organisms. (Conversely, our results also indicate that comparison of system-level features by the current methods is not feasible for the identification of precise phylogenetic relationships among species due to the fact that metabolic organization is intimately tied to the environment in which they evolved.) Also, the finding that Archaea and Eukarya proved related not only on informational-, but on metabolic grounds as well, is somewhat unexpected. While previous sequence comparisons identified informational genes of Eukaryotes being similar to those of Archaea, for operational (metabolic) genes bacterial and eukaryotic genes proved to be more closely related[6,18]. Eukaryotes thus apparently diverged from Archaea (see e.g., refs. 8, 9, 19), in-part as a result of robust horizontal transfer of operational genes[8,9] and/or perhaps as a by-product of (imperfect) phagotrophic consumption of Bacteria[20]. Yet, while transfer of physically-clustered, functionally-complementary bacterial genes into the archaean host's genome appears widespread[6,21], our analyses here demonstrate that the overall eukaryotic metabolic network architecture remained significantly less changed. This suggests that natural selection has favoured a selective retention of the bacterial enzymes that were presumably transferred into the proto-archaeal host's proteome.

What could be the underlying reason for this selectivity? We have previously shown that, irrespective of the particular pathways and species-specific enzymes utilized, the large-scale metabolic organization is essentially identical in all contemporary species, all possessing a robust and error-tolerant scale-free network architecture[11]. Also, alternative analytical approaches on the functional capabilities of metabolic networks indicate that the complete metabolic network is under organizational constraints[22,23]. Thus, we can infer a selective pressure during the symbiotic evolution of eukaryotes that limited their incorporation of bacterial metabolic enzymes, in order to maintain the already existing near optimal metabolic network architecture inherited from the proto-archaean host.

## Methods

**Database preparation**

For our analyses of metabolic and information transfer pathways we used the "Intermediate metabolism and Bioenergetics" and "Information transfer" portions of the WIT database (http://igweb.integratedgenomics.com/IGwit/)[10], respectively. This database predicts the existence of a biochemical pathway primarily based on the annotated genome of the organism combined with firmly established data from the biochemical literature. As of June 2000, this database provided description for 6 archaea (*A. pernix, A. fulgidus, M. thermoautotrophicum, M. jannaschii, P. furiosus, P. horikoshii*), 32 bacteria (*A. aeolicus, C. pneumoniae, C. trachomatis, Synechocystis sp., P. gingivalis, M. bovis, M. leprae, M. tuberculosis, B. subtilis, E. faecalis, C. acetobutylicum, M. genitalium, M. pneumoniae, S. pneumoniae, S. pyogenes, C. tepidum, R. capsulatus, R. prowazekii, N. gonorrhoeae, N. meningitidis, C. jejuni, H. pylori, E. coli, S. typhi, Y. pestis, A. mycetemcomitans, H. influenzae, P. aeruginosa, T. pallidum, B. burgdorferi, T. maritima, D. radiodurans*), and 5 eukaryotes (*E. nidulans, S. cerevisiae, C. elegans, O. sativa, A. thaliana*)(note the absence of metazoan Eukaryotes, including that of human, from the list). The downloaded data were manually rechecked, and synonyms and substrates without defined chemical identity were removed.

**Construction of network matrices and datasets**

Biochemical reactions described within a WIT pathway are composed of substrates and enzymes connected by directed links. For each reaction, educts and products were considered as nodes connected to the temporary educt-educt complexes and associated enzymes. For a given organism with $N$ substrates, $E$ enzymes and $R$ intermediate complexes the full stoichiometric interactions for metabolism and information transfer were compiled separately into an $(N+E+R) \times (N+E+R)$ matrix, generated separately for each of the 43 organisms. For a detailed description on the construction of network matrices, see ref. 11. The data sets, METAB/ENZ, METAB/SUBS, INFO/ENZ and INFO/SUBS were created with 834, 1267, 115 and 395 rows, respectively, all with 43 columns. In each matrix, score $x_{ij}$ indicates the number of times enzyme or substrate $i$ is present in the corresponding pathway of organism $j$.

**Enzyme and substrate ranking**

Enzymes and substrates present in the metabolic- and information transfer pathways of all 43 organisms were ranked based on the number of links each had in each organism, having considered incoming and outgoing links separately ($r=1$ was assigned to the enzyme and substrate with the largest number of connections, and $r=2$ to the second most connected one, and so on).

**Multivariate analyses**

Neighbor joining (NJ), unweighted group average clustering (UPGMA), ordinal clustering (OC) and non-metric multidimensional scaling (NMDS) were performed as previously described[12-15], and detailed in the Supplementary Information. All computations were made by the SYN-TAX 2000 software developed for WINDOWS systems[24].

**Analysis of the effect of database errors**

At the time of our analyses, of the 43 organisms whose metabolic- and information transfer network we have analyzed the genome of 25 has been completely sequenced (5 Archaea, 18 Bacteria, 2 Eukaryotes), while the remaining 18 are partially sequenced (i.e., the species underlined above). Therefore two major sources of possible errors in the database could affect our analysis: (a) the erroneous annotation of enzymes and consequently, biochemical reactions; for the organisms with completely sequenced genomes this is the likely source of error. (b) reactions and pathways missing from the database; for organisms with incompletely sequenced genomes both (a) and (b) are of potential source of error. We investigated the effect of database errors on the validity of our findings, the results being presented in the Supplementary Information indicating that the results offered in this paper are not affected by these errors.

**Acknowledgements**

We would like to acknowledge all members of the WIT project for making this invaluable database publicly available for the scientific community. Research at Eötvös University was supported by the Hungarian National Research Grant Foundation (OTKA), by the National Science Foundation at the University of Notre Dame, and at Northwestern University by grants from the National Cancer Institute.

**References**

1. Darwin, C. *The Origin of Species* (6th edition, J. Murray, London, 1872).
2. Maynard Smith, J. & Szathmáry, E. *The Major Transitions in Evolution* (Oxford University Press, Oxford, 1995).
3. Brooks, D.R. The nature of the organism. Life has a life of its own. *Ann. N. Y. Acad. Sci.* **901**, 257-265 (2000).
4. Woese, C.R., Kandler, O. & Wheelis, M.L. Towards a natural system of organisms: proposal for the domains Archaea, Bacteria, and Eucarya. *Proc. Nat. Acad. Sci. U.S.A.* **87**, 4576-4579 (1990).
5. Snel, B., Bork, P. & Huynen, M.A. Genome phylogeny based on gene content. *Nat. Genet.* **21**, 108-110 (1999).
6. Rivera, M.C., Jain, R., Moore, J.E. & Lake, J.A. Genomic evidence for two functionally distinct gene classes. *Proc. Nat. Acad. Sci. U.S.A.* **95**, 6239-6244 (1998).
7. Margulis, L. *Origin of eukaryotic cells* (Yale University Press, New Haven, CT, 1970).
8. Martin, W. & Müller, M. The hydrogen hypothesis for the first eukaryote. *Nature* **392**, 37-41 (1998).
9. Moreira, D. & Lopez-Garcia, P. Symbiosis between methanogenic archaea and delta-proteobacteria as the origin of eukaryotes: the syntrophic hypothesis. *J. Mol. Evol.* **47**, 517-530 (1998).
10. Overbeek, R. *et al.* WIT: integrated system for high-throughput genome sequence analysis and metabolic reconstruction. *Nucleic Acids Res.* **28**, 123-125 (2000).
11. Jeong, H., Tombor, B., Albert, R., Oltvai, Z.N. & Barabási, A.-L. The large-scale organization of metabolic networks. *Nature* **407**, 651-654 (2000).
12. Saitou, N. & Nei, M. The neighbor-joining method: a new method for reconstructing phylogenetic trees. *Mol. Biol. Evol.* **4**, 406-425 (1987).
13. Sokal, R. & Sneath, P. *Numerical Taxonomy*. (Freeman, San Francisco, 1973).
14. Podani, J. Explanatory variables in classifications and the detection of the optimum number of clusters. *In: C. Hayashi, N. Ohsumi, K. Yajima, Y. Tanaka, H.-H. Bock & Y. Baba (eds.), Data Science, Classification, and Related Methods*, Springer, Tokyo. pp. 125-132. (1998).
15. Podani, J. *Introduction to the Exploration of Multivariate Biological Data* (Backhuys, Leiden, 2000).
16. Tourasse, N.J. & Gouy, M. Accounting for evolutionary rate variation among sequence sites consistently changes universal phylogenies deduced from rRNA and protein-coding genes. *Mol. Phylogenet. Evol.* **13**, 159-168. (1999).
17. Andersson, J.O. & Andersson, S.G. Insights into the evolutionary process of genome degradation. *Curr. Opin. Genet. Dev.* **9**, 664-671 (1999).
18. Penny, D. & Poole, A. The nature of the last universal common ancestor. *Curr. Opin. Genet. Dev.* **9**, 672-677 (1999).
19. Miyata, T. *et al.* Evolution of archaebacteria: phylogenetic relationships among archaebacteria, eubacteria, and eukaryotes. *In: Owasa, S. & Honjo, T. (eds), Evolution of Life: Fossils, Molecules, and Culture,Springer-Verlag, Tokyo*, 337-351 (1991).
20. Doolittle, W.F. You are what you eat: a gene transfer ratchet could account for bacterial genes in eukaryotic nuclear genomes. *Trends Genet.* **14**, 307-311 (1998).
21. Lawrence, J.G. & Roth, J.R. Selfish operons: horizontal transfer may drive the evolution of gene clusters. *Genetics* **143**, 1843-1860 (1996).
22. Schilling, C.H., Letscher, D. & Palsson, B.O. Theory for the systemic definition of metabolic pathways and their use in interpreting metabolic function from a pathway-oriented perspective. *J. Theor. Biol.* **203**, 229-248 (2000).
23. Schuster, S., Fell, D.A. & Dandekar, T. A general definition of metabolic pathways useful for systematic organization and analysis of complex metabolic networks. *Nat. Biotechnol.* **18**, 326-332 (2000).
24. Podani, J. *SYN-TAX 2000. Computer Programs for Data Analysis in Ecology and Systematics* (Scientia Publishing, Budapest, 2001).

Correspondence and requests for materials should be addressed to Z.N.O. (zno008@northwestern.edu) or E. Sz.

(szathmary@zeus.colbud.hu)

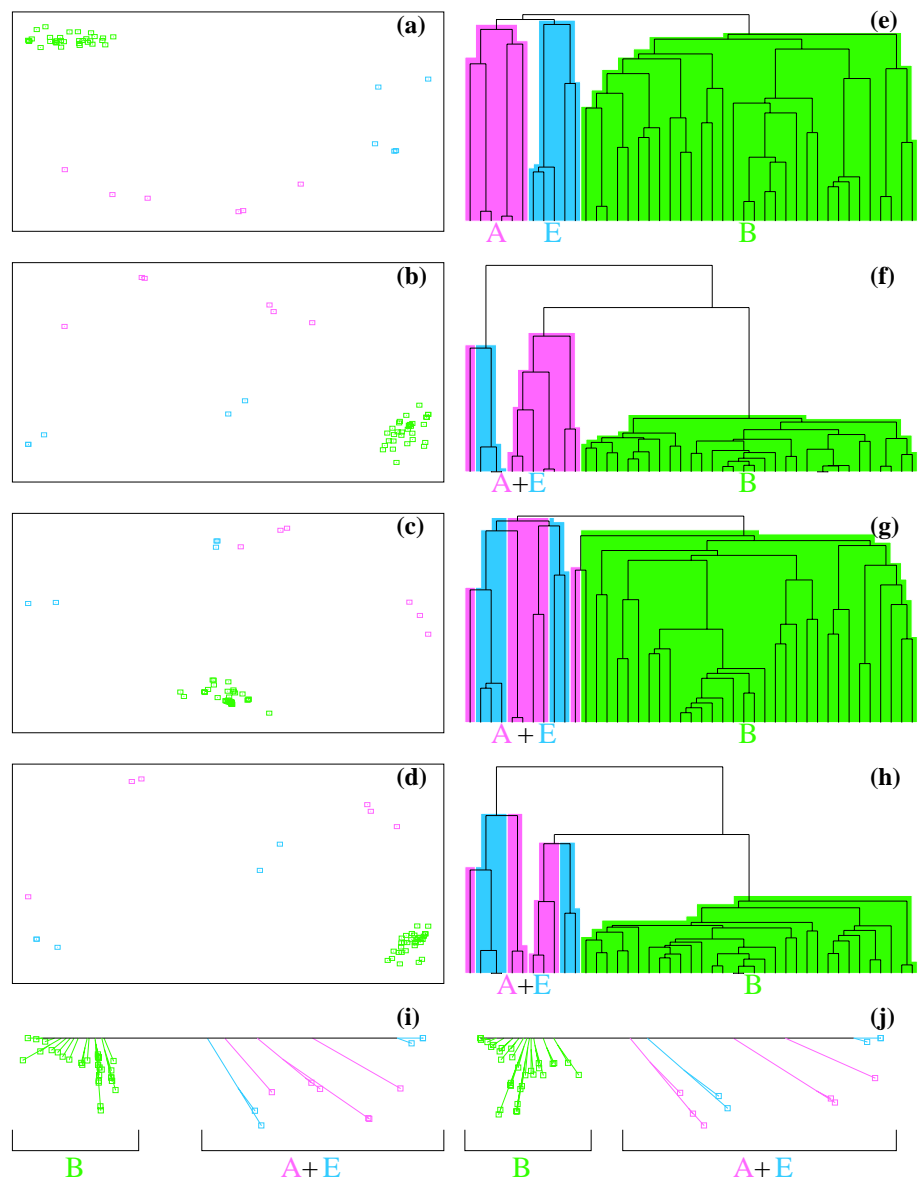

**Fig. 1**   Analyses based on Information transfer pathways

NMDS ordinations (*a-d*), OC (*e,g*) and UPGMA classifications (*f,h*), and unrooted NJ trees (*i,j*) for 43 taxa based on data on Information transfer pathways. (*a,b,e,f,i*) represent data based on substrate list, while (*c,d,g,h,j*) reflect on enzyme variables. (*a,c,e,g*) represent ordinal information, while (*b,d,f,h,i,j*) rely upon P/A information. Archaea (A) (magenta), Bacteria (B)(green) and Eukarya (E)(blue) are shown.

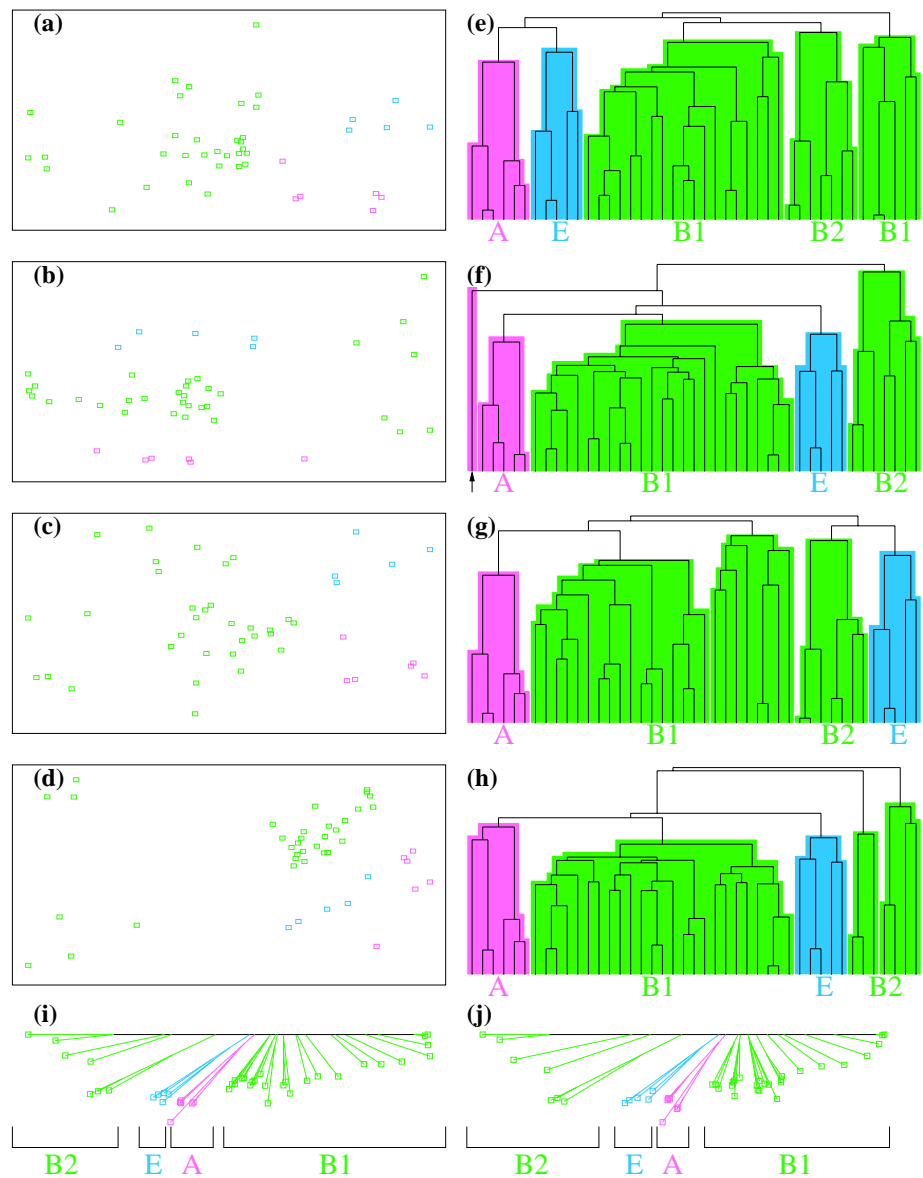

**Fig. 2**   Analyses based on Metabolic pathways

NMDS ordinations (*a-d*), OC (*e,g*) and UPGMA classifications (*f,h*), and unrooted NJ trees (*i,j*) for 43 taxa based on data on Metabolic pathways. (*a,b,e,f,i*) represent data based on substrate list, while (*c,d,g,h,j*) reflect on enzyme variables. (*a,c,e,g*) represent ordinal information, while (*b,d,f,h,i,j*) rely upon P/A information. Archaea (A) (magenta), Bacteria (B)(green) and Eukarya (E)(blue) are shown. The arrow in (*f*) denotes the Crenarchae, *A. pernix*.